\documentclass[useAMS,usenatbib]{mn2e}
\usepackage{natbib}
\usepackage{graphicx}
\usepackage{amssymb,latexsym,amsopn}
\usepackage[T1]{fontenc}

\newcommand{\da}{\partial}
\newcommand{\F}{\mathcal{F}}
\newcommand{\G}{\mathcal{G}}

\newcommand{\notag}{\nonumber}

\title[Shape, shear \& flexion]{Shape, shear \& flexion: An analytic flexion formalism for realistic mass profiles
\thanks{Research undertaken as part of the Commonwealth Cosmology Initiative (CCI:www.thecci.org), an international collaboration supported by the Australian Research Council}}
\author[P. D. Lasky and C. J. Fluke]{P. D. Lasky$^{1}$\thanks{E-mail: plasky@astro.swin.edu.au} and C. J. Fluke$^{1}$\thanks{E-mail: cfluke@astro.swin.edu.au}\\
$^{1}$Centre for Astrophysics \& Supercomputing, Swinburne University of Technology, PO Box 218, Hawthorn, Victoria, 3122, Australia}

\begin{document}

\pagerange{\pageref{firstpage}--\pageref{lastpage}}
\maketitle



\label{firstpage}

\begin{abstract}
Flexion is a non-linear gravitational lensing effect that arises from gradients in the convergence and shear across an image.  We derive a formalism that describes non-linear gravitational lensing by a circularly symmetric lens in the thin-lens approximation.  This provides us with relatively simple expressions for first- and second-flexion in terms of only the surface density and projected mass distribution of the lens.  We give details of exact lens models, in particular providing flexion calculations for a S\'ersic-law profile, which has become increasingly popular over recent years.  We further provide a single resource for the analytic forms of convergence, shear, first- and second-flexion for the following mass distributions: a point mass, singular isothermal sphere (SIS); Navarro-Frenk-White (NFW) profile; S\'ersic-law profile.  We quantitatively compare these mass distributions and show that the convergence and first-flexion are better indicators of the S\'ersic shape parameter, while for the concentration of NFW profiles the shear and second-flexion terms are preferred.   
\end{abstract}




\section{Introduction}
Quantifying the exact shape of baryonic and dark matter density profiles that form in expanding $\Lambda$CDM cosmologies is an ongoing issue.  N-body simulations suggest that CDM halos are well fitted by either Navarro-Frenk-White (NFW) density profiles \citep{navarro97} or deprojected {\it S\'ersic-like} laws in the form of \cite{einasto65} or \cite{prugniel97} density profiles \citep{navarro04,merritt05,merritt06,graham06,graham06a}.  Numerous observations of early-type galaxies suggest their luminosity profiles, and hence stellar mass distributions, follow either S\'ersic, core-S\'ersic or Nuker-law models \cite[e.g.][and references therein]{lauer95,lauer05,graham03a,ferrarese06a,cote06}, while gravitational lensing observations have suggested the {\it total} mass distribution (i.e. baryons plus dark matter) is consistently described by isothermal spheres \citep{treu02,treu04,rusin03,rusin05,koopmans06,gavazzi07,czoske08,dye08,tu09}.  Meanwhile, debate about the mass distributions of galaxy clusters has been no less intense; only recently is the NFW profile \cite[e.g.][]{carlberg97,vandermarel00,athreya02,katgert04,lin04,hansen05,lokas06,rines06,wojtak07,okabe09} being favoured over the isothermal sphere  \cite[e.g.][]{athreya02,ettori02,katgert04}.  As gravitational lensing traces total projected mass, it is an extremely powerful tool for determining the mass distributions of these systems without having to make assumptions about the dynamics or constitution of the lensing objects \cite[for recent reviews see][]{schneider05,hoekstra08}.

Traditionally, the study of weak lensing has been limited to linear effects; convergence and shear.  These fields have the effect that an elliptically shaped source galaxy gets mapped to an elliptical image.  Therefore, to determine information about the lensing object using first-order quantities, assumptions must be made about the intrinsic ellipticity of the source galaxy, or a large number of source galaxies must be utilised to ensure statistically reasonable results can be inferred \cite[see for e.g.][]{hoekstra04,mandelbaum06}.  Recently however, various authors have begun to consider higher-order lensing effects known as flexion \citep{goldberg02,goldberg05,bacon06}\footnote{\cite{irwin05,irwin06} also consider higher-order lensing, which they term {\it sextupole} lensing with components {\it sextupole}, {\it cardioid} and {\it displacement}.  In \cite{irwin07} the authors show that sextupole is equivalent to second-flexion and a combination of the cardioid and displacement terms is equivalent to first-flexion.}.  Flexion comes in two flavours that correspond to various spatial derivatives of the shear and convergence, implying they are due to gradients in the first-order fields across the extent of the source/image.  Physically, one can think of first-flexion as a shift in the centroid of the image with respect to the source and second-flexion as creating an arc-like structure in the image \citep{bacon06}.  That is, with the inclusion of flexion, an elliptical source galaxy gets mapped to a ``jelly-bean'' shaped image for a circularly symmetric lens.  In this way, flexion provides a better observable than the first-order fields as only one reasonable assumption about the source galaxy is required -- i.e. galaxies are not intrinsically flexed\footnote{Galaxies are not intrinsically flexed providing they are dynamically relaxed.}.

The flexion of a lensed image is formally calculated using multipole moments \citep{goldberg02,goldberg07,okura07,okura08}, however in this work we choose to treat the gravitational lensing variables as field variables.  That is, we determine the amount of convergence, shear and flexion one would measure as a function of the distance from the centre of the lensing mass and the angle in the sky, ignoring the overall shape and size of the source.  Determining the change in shape between the source/image pair requires a more detailed mapping that calculates small changes in the position of numerous light rays in the image and source planes.  This is a somewhat more difficult task that requires numerical methods, which is beyond the scope of the present work.  However, treating the gravitational lensing terms as field variables is extremely useful and the benefits it purveys are three-fold:
\begin{enumerate}
	\item It allows us to determine differences in the global gravitational lensing properties from various realistic density distributions (in particular see figure \ref{allallMpc} below).
	\item We can determine the relevant lensing terms (convergence, shear, first- or second-flexion) for discerning between individual shapes of density profiles.  For example in section \ref{comparesec} we show that convergence and first-flexion are good indicators of the S\'ersic shape parameter, whereas the concentration of NFW profiles can be determined by looking at the shear and second-flexion.
	\item The derivation of analytic solutions is a critical first step towards studying flexion through multiple lens planes with arbitrary mass distributions.  
\end{enumerate}

This paper is set out as follows;  In section \ref{secone} we systematically develop the two-dimensional thin-lens gravitational lens equation for an arbitrary, circularly symmetric matter distribution, deriving the first-order terms in section \ref{firstordersec} and the flexion terms in section \ref{flexionsec}.  In section \ref{exactsolnssec} we consider exact forms of the matter distributions, writing down analytic expressions for the convergence, shear and flexion for a point mass, singular isothermal sphere (SIS), NFW and S\'ersic profiles in sections \ref{Schwsec}, \ref{SISsec}, \ref{NFWsec} and \ref{Sersicsec} respectively.  In this way we are providing a single resource where the analytic forms for the convergence, shear, first- and second-flexion can be found for a range of useful density profiles.  Finally, in section \ref{comparesec} we compare the gravitational lensing effects of each of these profiles.  We find that the shear and second-flexion for a S\'ersic-law profile are systematically greater than for NFW and SIS profiles, whereas the convergence and first-flexion of each of the profiles are comparable.  We further show that the convergence and first-flexion provide excellent tracers for the S\'ersic shape parameter, whereas the shear and second-flexion are better indicators of the concentration parameter for the NFW profile.  We make some concluding remarks in section \ref{conc}.

\section{Analytic Lensing Formalism}\label{secone}
\subsection{The Thin-Lens Equation}\label{thinlenssec}
\begin{figure*}
\includegraphics[scale=0.4]{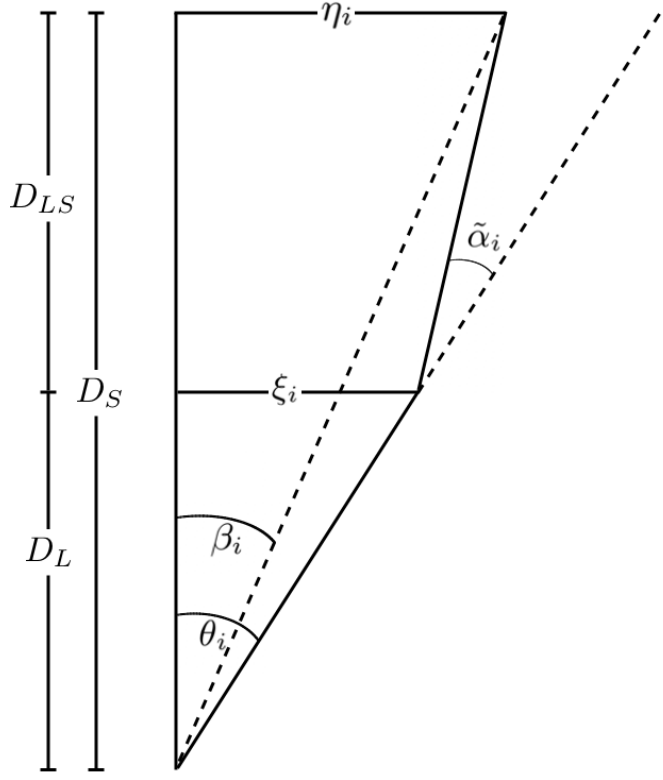}
\caption{Lens configuration.  Rather than use the angles, $\theta_{i}$ and $\beta_{i}$, to describe the position of the image and the source on the lens plane respectively, we use the two-dimensional radial coordinates denoted $\xi_{i}=D_{L}\theta_{i}$ and $\eta_{i}=D_{S}\beta_{i}$.  The distances between the source plane, lens plane and observer are all measured in terms of angular diameter distances, and $\tilde{\alpha}_{i}$ is the deflection angle.}
\label{Diag}
\end{figure*}

Given that flexion considers finite source sizes, we require a two-dimensional version of the thin-lens gravitational lens equation, whereby the mapping between a point on the source plane
and the image plane
are explicitly expressed.  In this way the thin-lens gravitational lens equation is expressed as 
\begin{equation}
	\eta_{i}=\frac{D_{S}}{D_{L}}\xi_{i}-D_{LS}\tilde{\alpha}_{i},\label{lens}
\end{equation}
where $\xi_{i}$ is the impact parameter on the image plane, $\eta_{i}$ is the distance between the origin of the coordinate system and the source on the source plane, $\tilde{\alpha}_{i}$ is the deflection angle and $D_{S}$, $D_{L}$ and $D_{LS}$ are the angular diameter distances from the observer to the source, the observer to the lens and the lens to the source respectively (the lens configuration is shown in figure \ref{Diag}).  Equation (\ref{lens}) can be put into a neater form by using angular coordinates, $\beta_{i}=\eta_{i}/D_{S}$ and $\theta_{i}=\xi_{i}/D_{L}$, such that
\begin{equation}
	\beta_{i}=\theta_{i}-\alpha_{i},\label{scalelens}
\end{equation}
where $\alpha_{i}=\tilde{\alpha}_{i}D_{LS}/D_{S}$ is the scaled deflection angle.  
Equation (\ref{scalelens}) can be re-expressed as a coordinate map between the two angular coordinate systems allowing the mapping to be expressed as a linear transformation
\begin{equation}
	\beta_{i}=A_{ij}\theta_{j},\label{linear}
\end{equation}
where $A_{ij}:=\da\beta_{i}/\da\theta_{j}$ is the Jacobian of transformation and summation is assumed over repeated indices.  Equation (\ref{linear}) implicitly assumes there are no gradients in the components of $A_{ij}$ across the image.  Physically, this implies that the convergence and shear are constant across the image, which is not necessarily a good approximation when the source has a finite spatial extent.  This is a reasonable assumption if sources are assumed to be point-like, however for realistic cosmology one would like to be able to probe deviations of the shear and convergence across images.  \cite{goldberg05} therefore generalized equation (\ref{linear}) to include higher-order terms, coined {\it flexion}.  These flexion terms account for an image's ``arciness'', and are therefore relevant if one is considering sources of finite spatial extent.  The non-linear expansion of the thin-lens equation is
\begin{equation}
	\beta_{i}=A_{ij}\theta_{j}+\frac{1}{2}D_{ijk}\theta_{j}\theta_{k},\label{secondorder}
\end{equation}
where 
\begin{equation}
	D_{ijk}:=\frac{\da A_{ij}}{\da \theta_{k}},
\end{equation}
is the gradient of the Jacobian.  The components of the $D_{ijk}$ matrix make up the components of the first- and second-flexion, a point that we discuss in considerably more detail in section \ref{flexionsec}.

We are interested in explicit expressions for the first- and second-order lensing terms with respect to positions on the image and source planes.  It is therefore instructive for us to work in the Cartesian $(\eta_{i},\xi_{i})$ system of coordinates rather than the angular $(\beta_{i},\theta_{i})$ coordinates (see figure \ref{Diag}).  In these coordinates, the Jacobian and its gradient are expressed as 
\begin{equation}
	A_{ij}=\frac{D_{L}}{D_{S}}\frac{\da\eta_{i}}{\da\xi_{j}}\qquad{\rm and}\qquad D_{ijk}=\frac{D_{L}^{2}}{D_{S}}\frac{\da^{2}\eta_{i}}{\da\xi_{j}\da\xi_{k}}.\label{6}
\end{equation}

In this article, we restrict our attention to distributions of matter which are circularly symmetric when projected on to the lens plane.  For such a profile, one can express the deflection angle as \cite[see for e.g.][]{schneider06}
\begin{equation}
	\tilde{\alpha}_{i}=\frac{4G}{c^{2}}\frac{M(|{\xi}|)}{|\xi|^{2}}\xi_{i},\label{alpha}
\end{equation}
where $|\xi|=\sqrt{\xi_{1}^{2}+\xi_{2}^{2}}$ and the projected mass, $M$, is defined as the area integral of the surface density, $\Sigma(|\xi|)$, which for circular symmetry is given by
\begin{equation}
	M(|\xi|)=2\pi\int_{0}^{|\xi|}\Sigma(\xi')\xi'{\rm d}\xi'.\label{massdef}
\end{equation}
Equation (\ref{lens}) can now be expressed in component form, also substituting equation (\ref{alpha}), yielding
\begin{equation}
	\eta_{i}=\xi_{i}\frac{D_{S}}{D_{L}}\left[1-\frac{1}{\pi\Sigma_{cr}}\frac{M(|\xi|)}{|\xi|^{2}}\right],\label{eta}
\end{equation}
where the critical surface density has been defined according to
\begin{equation}
	\Sigma_{cr}=\frac{c^{2}}{4\pi G}\frac{D_{S}}{D_{L}D_{LS}}.
\end{equation}

\subsection{First-Order Lensing}\label{firstordersec}
The first-order Jacobian, $A_{ij}$, can now be calculated by differentiating equation (\ref{eta}) which, after some algebra is
\begin{eqnarray}
	A_{11}&=&1-\frac{2\Sigma(|\xi|)\xi_{1}^{2}}{\Sigma_{cr}|\xi|^{2}}+\frac{M(|\xi|)}{\pi\Sigma_{cr}|\xi|^{4}}\left(\xi_{1}^{2}-\xi_{2}^{2}\right),\label{A11}\\
	A_{22}&=&1-\frac{2\Sigma(|\xi|)\xi_{2}^{2}}{\Sigma_{cr}|\xi|^{2}}+\frac{M(|\xi|)}{\pi\Sigma_{cr}|\xi|^{4}}\left(\xi_{2}^{2}-\xi_{1}^{2}\right),\label{A22}\\
	A_{12}&=&\frac{-2\xi_{1}\xi_{2}}{\pi\Sigma_{cr}|\xi|^{4}}{\cal Q}(|\xi|)=A_{21},\label{A12}
\end{eqnarray}
where for convenience we have defined the function
\begin{equation}
	{\cal Q}(\zeta):=\pi\Sigma(\zeta)\zeta^{2}-M(\zeta).
\end{equation}

The convergence and two components of the shear are expressed in terms of the components of the Jacobian as follows 
\begin{eqnarray}
	\gamma_{1}&=&-(A_{11}-A_{22})/2,\\
	\gamma_{2}&=&-A_{12},\\
	\kappa&=&1-(A_{11}+A_{22})/2.\label{kappa2}
\end{eqnarray}
Evaluating the convergence by substituting equations (\ref{A11}) and (\ref{A22}) into (\ref{kappa2}) implies
\begin{equation}
	\kappa=\frac{\Sigma(|\xi|)}{\Sigma_{cr}},\label{kappa}
\end{equation}
which is the familiar version of this equation \cite[see for e.g.][]{schneider06}.  Evaluating the two components of the shear using equations (\ref{A11}-\ref{A12}) implies
\begin{eqnarray}
	\gamma_{1}&=&\frac{{\cal Q}(|\xi|)}{\pi\Sigma_{cr}|\xi|^{4}}\left(\xi_{1}^{2}-\xi_{2}^{2}\right),\label{gamma1}\\
	\gamma_{2}&=&\frac{2{\cal Q}(|\xi|)}{\pi\Sigma_{cr}|\xi|^{4}}\xi_{1}\xi_{2}.\label{gamma2}
\end{eqnarray}
To compare the equations for the shear with a more familiar form, we convert the above equations from Cartesian coordinates, $(\xi_{1},\xi_{2})$, into polar coordinates on the lens plane, $(R,\phi)$, defined by $\xi_{1}=R\cos\phi$ and $\xi_{2}=R\sin\phi$.  This implies that the convergence and total shear, $\gamma=\gamma_{1}+i\gamma_{2}$, are
\begin{eqnarray}
	\kappa&=&\frac{\Sigma(R)}{\Sigma_{cr}},\label{kappaR}\\
	\gamma&=&\frac{{\cal Q}(R)}{\pi\Sigma_{cr}R^{2}}{\rm e}^{2i\phi}=|\gamma|{\rm exp}\left(2i\phi\right).\label{gammaR}
\end{eqnarray}
This is consistent with showing that the shear is a spin-two quantity, whereas the convergence is spin-zero.

Equations (\ref{kappaR}) and (\ref{gammaR}) give the convergence and shear distributions as a function of the mass profile of the galaxy.  Whilst the convergence is only dependent on the surface density, the shear is also a function of the mass.  This immediately gives the expected result for a point mass lens that the convergence is everywhere zero (except at the origin), whereas the shear is non-zero and goes as $M/R^{2}$ (see section \ref{Schwsec}).

\subsection{Flexion}\label{flexionsec}
The real interest in this article lies in the flexion terms, which are given by the second-order Taylor series expansion, equation (\ref{secondorder}).  The three-tensor, $D_{ijk}$, is expressed in terms of the gradient of the linear Jacobian of transformation in equation (\ref{6}) that, 
given equations (\ref{A11}-\ref{A12}), allows us to calculate the individual components of the three-tensor in terms of the coordinates $\xi_{1}$ and $\xi_{2}$.  After much algebra, one finds
\begin{eqnarray}
	D_{111}&=&\frac{2D_{L}\xi_{1}\left(\xi_{1}^{2}-3\xi_{2}^{2}\right)}{\pi\Sigma_{cr}|\xi|^{6}}{\cal Q}-\frac{2D_{L}\xi_{1}^{3}}{\Sigma_{cr}|\xi|^{3}}\frac{d\Sigma}{d|\xi|},\label{D111}\\
	D_{211}&=&\frac{2D_{L}\xi_{2}\left(3\xi_{1}^{2}-\xi_{2}^{2}\right)}{\pi\Sigma_{cr}|\xi|^{6}}{\cal Q}-\frac{2D_{L}\xi_{1}^{2}\xi_{2}}{\Sigma_{cr}|\xi|^{3}}\frac{d\Sigma}{d|\xi|},\label{D211}\\
	D_{221}&=&\frac{-2D_{L}\xi_{1}\left(\xi_{1}^{2}-3\xi_{2}^{2}\right)}{\pi\Sigma_{cr}|\xi|^{6}}{\cal Q}-\frac{2D_{L}\xi_{1}\xi_{2}^{2}}{\Sigma_{cr}|\xi|^{3}}\frac{d\Sigma}{d|\xi|},\label{D221}\\
	D_{222}&=&\frac{-2D_{L}\xi_{2}\left(3\xi_{1}^{2}-\xi_{2}^{2}\right)}{\pi\Sigma_{cr}|\xi|^{6}}{\cal Q}-\frac{2D_{L}\xi_{2}^{3}}{\Sigma_{cr}|\xi|^{3}}\frac{d\Sigma}{d|\xi|},\label{D222}
\end{eqnarray}
where for the remainder of the article $\Sigma=\Sigma(|\xi|)$, ${\cal Q}={\cal Q}(|\xi|)$ and $M=M(|\xi|)$ unless otherwise explicitly stated.
As shown by \cite{bacon06}, the three-tensor, $D_{ijk}$, can be expressed as the sum of two other tensors, $D_{ijk}=\F_{ijk}+\G_{ijk}$, which can be written component-wise as
\begin{eqnarray}
	\F_{ij1}&=&-\frac{1}{2}\left(\begin{array}{cc} 
		3\F_{1} & \F_{2} \\
		\F_{2} & \F_{1}\end{array}\right),\\
	\F_{ij2}&=&-\frac{1}{2}\left(\begin{array}{cc} 
		\F_{2} & \F_{1} \\
		\F_{1} & 3\F_{2}\end{array}\right),\\
	\G_{ij1}&=&-\frac{1}{2}\left(\begin{array}{cc} 
		\G_{1} & \G_{2} \\
		\G_{2} & -\G_{1}\end{array}\right),\\
	\G_{ij2}&=&-\frac{1}{2}\left(\begin{array}{cc} 
		\G_{2} & -\G_{1} \\
		-\G_{1} & -\G_{2}\end{array}\right).
\end{eqnarray}
In the above, $\F=\F_{1}+i\F_{2}$ and $\G=\G_{1}+i\G_{2}$ are known as first- and second-flexion respectively \citep{bacon06}.  First-flexion is a spin-one quantity that measures the shift in the centroid of the image, and second-flexion is a spin-three quantity measuring the ``arciness'' of the image.

Inverting the above system of equations implies we can express the components of the first- and second-flexion in terms of the components of $D_{ijk}$;
\begin{eqnarray}
	\F_{1}&=&-\frac{1}{2}(D_{111}+D_{221}),\label{F1}\\
	\F_{2}&=&-\frac{1}{2}(D_{211}+D_{222}),\label{F2}\\
	\G_{1}&=&-\frac{1}{2}(D_{111}-3D_{221}),\label{G1}\\
	\G_{2}&=&-\frac{1}{2}(3D_{211}-D_{222}).\label{G2}
\end{eqnarray}
Therefore, utilizing equations (\ref{D111}-\ref{D222}), (\ref{F1}) and (\ref{F2}), one can show that the components of the first-flexion are 
\begin{eqnarray}
	\F_{1}&
	=&\frac{D_{L}}{\Sigma_{cr}}\frac{\da\Sigma}{\da\xi_{1}},\label{F1gen}\\
	\F_{2}&
	=&\frac{D_{L}}{\Sigma_{cr}}\frac{\da\Sigma}{\da\xi_{2}}.\label{F2gen}
\end{eqnarray}
That is, the first and second components of first-flexion (i.e., $\F_{1}$ and $\F_{2}$ respectively) are the directional derivatives of the surface density.  Note that the first-flexion terms do not include any functions of the mass, which is again consistent with the flexion due to a point mass being zero (see section \ref{Schwsec}).  The total first-flexion, $\F=\F_{1}+i\F_{2}$, is given by
\begin{eqnarray}
	\F&=&\frac{D_{L}}{\Sigma_{cr}|\xi|}\frac{d\Sigma}{d|\xi|}\left(\xi_{1}+i\xi_{2}\right)\notag\\
		&=&\frac{D_{L}}{\Sigma_{cr}}\frac{d\Sigma(R)}{dR}{\rm e}^{i\phi},\label{1flexionR}
\end{eqnarray}
implying the first-flexion is a spin-one field \citep{bacon06} which is the gradient of the surface density.  Given that the convergence is proportional to the surface density, this is an equivalent way of saying that the first-flexion is the gradient of the convergence.

Following the same procedure outlined above, one can show that the components of the second-flexion are
\begin{eqnarray}
	\G_{1}&=&\frac{D_{L}\xi_{1}\left(\xi_{1}^{2}-3\xi_{2}^{2}\right)}{\pi\Sigma_{cr}|\xi|^{6}}\left[\pi\frac{d\Sigma}{d|\xi|}|\xi|^{3}-4{\cal Q}\right],\\
	\G_{2}&=&\frac{D_{L}\xi_{2}\left(3\xi_{1}^{2}-\xi_{2}^{2}\right)}{\pi\Sigma_{cr}|\xi|^{6}}\left[\pi\frac{d\Sigma}{d|\xi|}|\xi|^{3}-4{\cal Q}\right].
\end{eqnarray}
These form the total second-flexion term, $\G=\G_{1}+i\G_{2}$, as
\begin{equation}
	\G=\frac{D_{L}}{\pi\Sigma_{cr}|\xi|^{6}}\left[\pi\frac{d\Sigma}{d|\xi|}|\xi|^{3}-4{\cal Q}\right]\left(\xi_{1}+i\xi_{2}\right)^{3}.
\end{equation}
Finally, expressing the second-flexion in polar coordinates and expanding ${\cal Q}$ to show the explicit dependence on the surface mass density and the projected mass one finds
\begin{equation}
	\G=\frac{D_{L}}{\Sigma_{cr}}\left[\frac{d\Sigma(R)}{dR}-\frac{4\Sigma(R)}{R}+\frac{4M(R)}{\pi R^{3}}\right]{\rm e}^{3i\phi}.\label{2flexionR}
\end{equation}
This expression is relatively simple and shows the dependence of second-flexion on the surface density, its divergence and the projected mass distribution.  Moreover, one can explicitly see that this is a spin-three field, as shown in \cite{bacon06}.

It is pertinent to note that \cite{schneider08} showed that first- and second-flexion are not observable quantities due to the mass-sheet degeneracy.  Instead, they showed that the observable spin-one and spin-three fields are the {\it reduced flexion} terms;
\begin{equation}
	\frac{\F+g\F^{*}}{1-\kappa}\qquad{\rm and}\qquad\frac{\G+g\F}{1-\kappa},\label{42}
\end{equation}	
where $g=\gamma/(1-\kappa)$ is the reduced shear and $\F^{*}=\F_{1}-i\F_{2}$ is the complex conjugate of first-flexion.  Whilst it is beneficial to keep the mass-sheet degeneracy in mind, a majority of work on measuring flexion in real images using either the shapelets \citep{refregier03a,refregier03b,goldberg05} or HOLICs \citep{okura07,okura08,goldberg07} approaches have been based on decomposing the higher-order components of the images into $\F$ and $\G$.  Therefore, for this article we continue to focus on the flexion quantities, $\F$ and $\G$, rather than the terms expressed in (\ref{42}).

\subsection{A Note On Units}
The system of coordinates we are using in this article are somewhat unconventional in the sense that they use distance, rather than angular coordinates on the image and source planes.  We do this as we believe these units are more conducive to numerical applications, and are also more descriptive to the reader.  However, our approach implies it is worth spending some time elucidating the dimensions that we are using for each of the derived quantities.

The surface density is defined as the projection of the three-dimensional density distribution, implying it has units of mass per unit area.  Integrating this according to equation (\ref{massdef}) implies the two-dimensional (projected) mass distribution has units of mass (as one would expect).  Therefore, the convergence and shear given by equations (\ref{kappa}), (\ref{gamma1}) and (\ref{gamma2}) are dimensionless.  This is not surprising, and indeed this is true when one uses angular coordinates.  The flexion terms presented here are in dimensions of (angle)$^{-1}$, which is also consistent with those expressed in angular coordinates, however the angle is in units of radians.  This can be seen as the expressions for the first- and second-flexion, equations (\ref{1flexionR}) and (\ref{2flexionR}) respectively, are proportional to the ratio of the angular diameter distance from the observer to the lens plane, $D_{L}$, and the impact parameter, $|\xi|$.  As discussed in section \ref{thinlenssec}, this ratio is $\theta^{-1}$.  This implies that to compare the results presented in this paper with those derived using angular coordinates \citep[for example][]{bacon06}, one must convert from units of radians to arcseconds.  As an alternative, in appendix \ref{anglesapp} we show the equations for convergence, shear and flexion expressed in angular coordinates.

\section{Exact Solutions}\label{exactsolnssec}
In this section we present analytic solutions of the equations expressed hitherto for various circularly symmetric matter distributions.  For completeness, we first discuss the simplest case of a point mass, then move on to SIS, NFW and finally S\'ersic-law profiles.  These exact solutions are then used in section \ref{comparesec} to investigate differences between the flexion of various lenses, and we discuss how these flexion terms can be used to constrain the density distributions.  

\subsection{Point Mass}\label{Schwsec}
Consider a Schwarzschild lens (i.e. a point mass) situated at the origin of the coordinate system on the lens plane.  This implies that $\Sigma=\delta(|\xi|)$, where $\delta$ is the Dirac delta function.  Moreover, the projected mass distribution $M=M_{s}$ is a constant for all $|\xi|\neq0$.  The Einstein radius for such a system in angular coordinates is 
\begin{equation}
	\theta_{E}=\left(\frac{4GM_{s}}{c^{2}}\frac{D_{LS}}{D_{S}D_{L}}\right)^{1/2},
\end{equation}
and the critical surface density and Einstein radius are related according to 
\begin{equation}
	\Sigma_{cr}=\frac{M_{s}}{\pi D_{L}^{2}\theta_{E}^{2}}=\frac{M_{s}}{\pi\xi_{E}^{2}},
\end{equation}
where we have defined $\xi_{E}:=D_{L}\theta_{E}$ to be the Einstein radius measured as a distance on the lens plane.  

The convergence is linearly proportional to the surface density of the system, equation (\ref{kappa}), implying it is trivially zero everywhere except $|\xi|=0$.  The shear, however, includes terms involving the projected mass of the system.  From equations (\ref{gamma1}) and (\ref{gamma2}) the first and second components of the shear are 
\begin{equation}
	\gamma_{1}=\frac{\xi_{E}^{2}}{|\xi|^{4}}\left(\xi_{2}^{2}-\xi_{1}^{2}\right)\qquad{\rm and}\qquad\gamma_{2}=\frac{-2\xi_{E}^{2}}{|\xi|^{4}}\xi_{1}\xi_{2},
\end{equation}
and the total shear is
\begin{equation}
	\gamma=\frac{-\xi_{E}^{2}}{|\xi|^{4}}\left(\xi_{1}+i\xi_{2}\right)^{2}=\frac{-\theta_{E}^{2}}{\theta^{2}}{\rm e}^{2i\phi}.\label{schgamma}
\end{equation}
Note that the final equation in (\ref{schgamma}) is the usual expression for the shear associated with a point mass, however with an additional negative sign.  This is a direct result of the calculation, however we note that the negative sign is due to the choice of coordinates.  That is, rotating our coordinate system by ninety degrees implies the negative sign vanishes [i.e. $\phi\rightarrow\phi+\pi/2$ implies $\exp(2i\phi)\rightarrow-\exp(2i\phi)$].  Therefore, one is free to scale away the negative sign in equation (\ref{schgamma}) by rotating the coordinate system, which yields the familiar result for the shear induced by a point mass lens.


According to equations (\ref{F1gen}) and (\ref{F2gen}), the components of the first-flexion are the directional derivatives of the surface density, implying these vanish (for all $|\xi|\neq0$);
\begin{equation}
	\F_{1}=\F_{2}=\F=0.
\end{equation}
As mentioned above, first-flexion is associated with a shift of the centroid of the image with respect to the source.  Therefore, $\F\equiv0$ for a point mass implies the centroid of the image is unchanged.  Second-flexion, however, is associated with the ``arciness'' of the image, and one finds for the point mass lens that
\begin{eqnarray}
	\G_{1}&=&\frac{4D_{L}\xi_{E}^{2}}{|\xi|^{6}}\xi_{1}\left(\xi_{1}^{2}-3\xi_{2}^{2}\right),\\
	\G_{2}&=&\frac{4D_{L}\xi_{E}^{2}}{|\xi|^{6}}\xi_{2}\left(3\xi_{1}^{2}-\xi_{2}^{2}\right).
\end{eqnarray}
Combining these, the total second-flexion is given by
\begin{equation}
	\G=\frac{4D_{L}\xi_{E}^{2}}{|\xi|^{6}}\left(\xi_{1}+i\xi_{2}\right)^{3}=\frac{4\theta_{E}^{2}}{\theta^{3}}{\rm e}^{3i\phi}.
\end{equation}
Taking the magnitude of the above expression implies second-flexion decreases proportionally to the projected radius cubed, $|\G|\propto|\xi|^{-3}$, compared to the shear which decreases proportionally to the projected radius squared, $|\gamma|\propto|\xi|^{-2}$.  Therefore, as one gets further from the source, the shear term will come to dominate over the second-flexion, and this effect will become more negligible as one looks further from the origin (see figure \ref{allallMpc}).  

It is interesting to note that while the point mass has vanishing first-flexion and non-zero second-flexion, it is relatively straightforward to also construct lens models where the second-flexion vanishes but the first-flexion is non-zero.  Indeed by setting equation (\ref{2flexionR}) equal to zero, one can show that the second-flexion vanishes for models with $\Sigma\propto |\xi|^{2}$.  Whilst this is obviously unphysical, as the surface density distribution increases monotonically as a function of radius, one still finds that the first-flexion is non-zero everywhere (for all $|\xi|\neq0$) with $|\F|\propto|\xi|$.

\subsection{Singular Isothermal Sphere}\label{SISsec}
Several observational studies of early-type galaxies using gravitational lensing have suggested that the total matter distribution (i.e. baryonic plus dark matter) is well described by a nearly isothermal density profile \citep{treu02,treu04,rusin03,rusin05,koopmans06,czoske08,dye08,tu09}.  Indeed the strongest evidence for this is from \cite{gavazzi07} who reported weak lensing studies of 22 early-type galaxies based on $HST$ imaging, concluding that an isothermal distribution is consistent out to 100 effective radii.  On larger scales, isothermal spheres have been found to fit the density profile of galaxy clusters \citep{athreya02,ettori02,katgert04}, although recent debate has suggested the NFW profile may provide a better fit (see section \ref{NFWsec}).  Analytically, gravitational lensing studies have compared SIS and NFW profiles at first- \citep{wright00} and higher-order \citep{bacon06}, implying the SIS profile provides us with the simplest realistic profile to reproduce known results.

The SIS density profile is described in terms of the three-dimensional (deprojected) radius, $r$, as
\begin{equation}
	\rho(r)=\frac{\sigma^{2}}{2\pi Gr^{2}},\label{SISrho}
\end{equation}
where $\sigma$ is the one-dimensional velocity dispersion.  Projecting equation (\ref{SISrho}) along the line-of-sight gives the surface density \citep[for example see][]{binney87}
\begin{equation}
	\Sigma=\frac{\sigma^{2}}{2G|\xi|}.
\end{equation}
We note that the SIS profile, like the point mass, is singular at the origin.  This implies that the following analysis is valid for all $|\xi|\neq0$, however in reality one must quantify at what radius the system is a ``weak lens'' (see section \ref{comparesec}).  
By defining the Einstein radius in units of distance as $\xi_{E}:=D_{L}\theta_{E}$, where $\theta_{E}$ is the standard Einstein radius for an SIS lens, one can show that the critical surface density, $\Sigma_{cr}$, is related to the Einstein radius as
\begin{equation}
	\Sigma_{cr}=\frac{\sigma^{2}}{G\xi_{E}}.
\end{equation}
By substituting the above equations into (\ref{kappa}), we can show that the convergence falls-off linearly with the distance from the center of the coordinate system;
\begin{equation}
	\kappa=\frac{\xi_{E}}{2|\xi|}=\frac{\theta_{E}}{2\theta}.
\end{equation}
Moreover, the components of the shear have a similar behaviour with the extra angular dependence
\begin{eqnarray}
	\gamma_{1}&=&\frac{-\xi_{E}}{2|\xi|^{3}}\left(\xi_{1}^{2}-\xi_{2}^{2}\right),\\
	\gamma_{2}&=&\frac{-\xi_{E}}{|\xi|^{3}}\xi_{1}\xi_{2}.
\end{eqnarray}
As we know the shear is a spin-two field, we are essentially only interested in the strength of the respective fields.  Therefore, for the remainder of the article, unless explicitly stated, we shall just be working with the magnitudes of these fields.  Now, as $\gamma=|\gamma|\exp(2i\phi)$, one can show from the above that
\begin{equation}
	|\gamma|=\frac{\xi_{E}}{2|\xi|}=\frac{\theta_{E}}{2\theta}.
\end{equation}
The first-flexion components are
\begin{equation}
	\F_{1}=\frac{-D_{L}\xi_{E}}{2|\xi|^{3}}\xi_{1}\qquad{\rm and}\qquad
	\F_{2}=\frac{-D_{L}\xi_{E}}{2|\xi|^{3}}\xi_{2}.\label{SISfirsttwo}
\end{equation}
First-flexion is a spin-one field, implying $\F=|\F|\exp(i\phi)$.  Equations (\ref{SISfirsttwo}) imply the first-flexion falls-off proportionally to the distance squared;
\begin{equation}
	|\F|=\frac{-D_{L}\xi_{E}}{2|\xi|^{2}}=\frac{-\theta_{E}}{2\theta^{2}}.
\end{equation}
The second-flexion components are
\begin{eqnarray}
	\G_{1}&=&\frac{3D_{L}\xi_{E}}{2|\xi|^{5}}\xi_{1}\left(\xi_{1}^{2}-3\xi_{2}^{2}\right),\\
	\G_{2}&=&\frac{3D_{L}\xi_{E}}{2|\xi|^{5}}\xi_{2}\left(3\xi_{1}^{2}-\xi_{2}^{2}\right),
\end{eqnarray}
which is a spin-three field, $\G=|\G|\exp(3i\phi)$, implying second-flexion also decreases proportionally to the radius squared
\begin{equation}
	|\G|=\frac{3D_{L}\xi_{E}}{2|\xi|^{2}}=\frac{3\theta_{E}}{2\theta^{2}}.
\end{equation}
All of the above equations that are written in angular coordinates are consistent with those presented in \cite{bacon06}.  These analytic expressions are plotted for a specific mass in figure \ref{allallMpc} where they are compared with the NFW and S\'ersic-law profiles.

\subsection{Navarro-Frenk-White Profile}\label{NFWsec}
Compared to the SIS profile, the NFW profile introduces an extra parameter into the scaling of density distributions.  The concentration, $c$, which is defined as the ratio of the three-dimensional virial radius to the three-dimensional scale radius, $c=r_{\Delta}/r_{s}$, is a function of the particular cosmology being used.  In section \ref{comparesec} we compare the NFW profile to the other profiles being analysed here, as well as comparing the effects of varying the concentration on the first- and higher-order lensing phenomena.  The NFW profile is generally given in terms of the three-dimensional radii, $r$,
\begin{equation}
	\rho(r)=\frac{\delta_{c}\rho_{c}}{\left(r/r_{s}\right)\left(1+r/r_{s}\right)^{2}},
\end{equation}
where 
$\rho_{c}$ is the critical density of the Universe and
\begin{equation}
	\delta_{c}=\frac{\Delta}{3}\frac{c^{3}}{\ln\left(1+c\right)-c/\left(1+c\right)}.
\end{equation}
Projecting this onto the two-dimensional radius, $|\xi|$, gives the surface density \citep{bartelmann96}
\begin{equation}
	\Sigma=\frac{2\rho_{c}\delta_{c}r_{s}^{3}}{|\xi|^{2}-r_{s}^{2}}\left[1-\Xi(|\xi|)\right],\label{surfaceNFW}
\end{equation}
where we have defined the following function
\begin{equation}
	\Xi(|\xi|):=\left\{
		\begin{array}{lc}
			\frac{2r_{s}}{\sqrt{r_{s}^{2}-|\xi|^{2}}}{\rm arctanh}\sqrt{\frac{r_{s}-|\xi|}{r_{s}+|\xi|}} &|\xi|<r_{s}\\
			\frac{2r_{s}}{\sqrt{|\xi|^{2}-r_{s}^{2}}}{\rm arctan}\sqrt{\frac{|\xi|-r_{s}}{|\xi|+r_{s}}} & |\xi|>r_{s}
		\end{array}\right..\label{fdef}
\end{equation}
Integrating the surface density gives the projected mass distribution
\begin{equation}
	M=4\pi\rho_{c}\delta_{c}r_{s}^{3}\left[\ln\frac{|\xi|}{2r_{s}}+\Xi(|\xi|)\right].
\end{equation}


The convergence and shear can now be expressed simply as functions of the above expressions.  From equation (\ref{kappaR}), the convergence is simply $\kappa=\Sigma/\Sigma_{cr}$, where the surface density is given by equation (\ref{surfaceNFW}).  The total shear is determined by equation (\ref{gammaR}) as 
\begin{equation}
	|\gamma|=\frac{2\rho_{c}\delta_{c}r_{s}^{3}}{\Sigma_{cr}\left(|\xi|^{2}-r_{s}^{2}\right)}\left[1-\Xi-2\left(1-\frac{r_{s}^{2}}{|\xi|^{2}}\right)\left(\ln\frac{|\xi|}{2r_{s}}+\Xi\right)\right],
\end{equation}
where $\Xi=\Xi(|\xi|)$.  Equation (\ref{1flexionR}) implies first-flexion is found by differentiating the surface density, which can be shown to be
\begin{equation}
	|\F|=\frac{-2D_{L}\rho_{c}\delta_{c}r_{s}^{3}}{\Sigma_{cr}|\xi|\left(|\xi|^{2}-r_{s}^{2}\right)^{2}}\left(2|\xi|^{2}+r_{s}^{2}-3|\xi|^{2}\Xi\right),
\end{equation}
and equation (\ref{2flexionR}) implies the second-flexion for the NFW profile is given by the expression
\begin{eqnarray}
	|\G|&=&\frac{2D_{L}\rho_{c}\delta_{c}r_{s}^{3}}{\Sigma_{cr}|\xi|\left(|\xi|^{2}-r_{s}^{2}\right)^{2
}}\Bigg[8\left(1-\frac{r_{s}^{2}}{|\xi|}\right)^{2}\ln\frac{|\xi|}{2r^{s}}\notag\\
	&&+3\left(r_{s}^{2}-2|\xi|^{2}\right)+\left(15|\xi|^{2}-20r_{s}^{2}+8\frac{r_{s}^{4}}{|\xi|^{2}}\right)\Xi\Bigg]
\end{eqnarray}
Despite the mass of the system being infinite, one can show that the convergence, shear, first- and second-flexion all tend to zero as $|\xi|\rightarrow\infty$.  These profiles are plotted against the SIS and S\'ersic-law profiles in figure \ref{allallMpc}, and we also look at the dependence of the concentration in figure \ref{NFWvarycall} and section \ref{comparesec}.

\subsection{S\'ersic Profile}\label{Sersicsec}
It has long been argued that a S\'ersic-law \citep{sersic68} provides a remarkably good fit to luminosity profiles of early-type galaxies, ranging in size from dwarf galaxies to the largest elliptical galaxies \citep[][]{caon93,graham01,graham03,graham03a,trujillo04}.  For a concise reference to S\'ersic quantities, see \cite{graham05}.  Recently, a S\'ersic-law has also been shown to provide a good fit to three-dimensional density profiles \citep{navarro04}, and also to projected surface density profiles \citep{merritt05} of dark matter halos.  In a series of papers \citep{merritt06,graham06,graham06a} it has further been shown that projected S\'ersic surface density profiles provide the best fit to simulated galaxy- and cluster-sized dark matter halos. 

\cite{cardone04} first analysed the S\'ersic profile in the gravitational lensing context, showing that mass estimates using lens reconstructions is highly dependent on the choice of S\'ersic parameter.  \cite{eliasdottir07} compared gravitational lensing for S\'ersic and NFW profiles, and found that mass estimates may differ by up to a factor of two, dependent on the choice of density profile and S\'ersic index.  In the weak lensing regime, they did this by looking at the shear of both profiles.  We take this a step further by also analysing higher-order lensing terms.  In this section, we provide the first explicit representation of flexion terms for the S\'ersic-law profile, and in section \ref{comparesec} we compare these results to those of the NFW and SIS profiles. 

The S\'ersic profile is defined in terms of the surface density
\begin{equation}
	\ln\left(\frac{\Sigma}{\Sigma_{e}}\right)=-b_{n}\left[\left(\frac{|\xi|}{\xi_{e}}\right)^{1/n}-1\right],\label{Sersic}
\end{equation}
where $\Sigma_{e}$ is the surface density at the effective radius, $\xi_{e}$.  The constant $n$ is the S\'ersic shape parameter which describes the shape of the profile and $b_{n}$ is a function of $n$ that is chosen such that the effective radius contains half of the projected mass of the system.  Analytically, this is given as the solution of $\Gamma(2n)=2\Gamma(2n,\,b_{n})$, where 
\begin{equation}
	\Gamma(\alpha,\,x)=\int_{t=0}^{x}{\rm e}^{-t}t^{\alpha-1}dt,
\end{equation}
is the lower incomplete gamma function and $\Gamma(\alpha):=\lim_{x\rightarrow\infty}\Gamma(\alpha,\,x)$ is the complete gamma function.  As such, $b_{n}$ can be reasonably approximated to $b_{n}=2n-1/3+4/(405n)+{\cal O}(n^{-2})$ for $0.5<n<10$ \citep{ciotti99}.  Integrating the surface density gives the projected mass
\begin{equation}
	M=2\pi n\frac{{\rm e}^{b_{n}}}{b_{n}^{2n}}\Sigma_{e}\xi_{e}^{2}\Gamma\left(2n,\,Z\right),
\end{equation}
where $Z=b_{n}\left(|\xi|/\xi_{e}\right)^{1/n}$.  One can see that the S\'ersic-law has one more parameter than the NFW profile, a point we discuss in more detail in section \ref{comparesec} and also appendix \ref{PSapp}.

After much algebra, one can show that the magnitudes of the convergence and shear for the S\'ersic profile can be expressed as
\begin{eqnarray}
	\kappa&=&\frac{\Sigma_{e}}{\Sigma_{cr}}\exp\left\{b_{n}\left[1-\left(\frac{|\xi|}{\xi_{e}}\right)^{1/n}\right]\right\},\\
	|\gamma|&=&\frac{\Sigma_{e}}{\Sigma_{cr}}\Bigg(\exp\left\{b_{n}\left[1-\left(\frac{|\xi|}{\xi_{e}}\right)^{1/n}\right]\right\}\notag\\
	&&-\frac{2n{\rm e}^{b_{n}}\xi_{e}^{2}}{b_{n}^{2n}{|\xi|^{2}}}\Gamma\left(2n,\,Z\right)\Bigg).
\end{eqnarray}
Furthermore, the magnitudes of the first- and second-flexion terms are
\begin{eqnarray}
	|\F|&=&\frac{-D_{L}\Sigma_{e}b_{n}}{n\Sigma_{cr}|\xi|}\left(\frac{|\xi|}{\xi_{e}}\right)^{1/n}\exp\left\{b_{n}\left[1-\left(\frac{|\xi|}{\xi_{e}}\right)^{1/n}\right]\right\},\\
	|\G|&=&\frac{D_{L}\Sigma_{e}}{\Sigma_{cr}}\Bigg(\frac{8n{\rm e}^{b_{n}}\xi_{e}^{2}}{b_{n}^{2n}|\xi|^{3}}\Gamma\left(2n,\,Z\right)-\frac{1}{|\xi|}\Bigg[\frac{b_{n}}{n}\left(\frac{|\xi|}{\xi_{e}}\right)^{1/n}-4\Bigg]\notag\\
	&&\times\exp\left\{b_{n}\left[1-\left(\frac{|\xi|}{\xi_{e}}\right)^{1/n}\right]\right\}\Bigg).
\end{eqnarray}

\section{Profile Comparisons}\label{comparesec}
We compare gravitational lensing effects of the various density distributions by holding the virial mass of each system constant, which is the same method used by \cite{wright00} for comparing first-order lensing properties of the NFW and SIS profiles.  The three-dimensional (i.e. deprojected) virial radius, $r_{\Delta}$, is defined as the radius inside which the average density of the halo is $\Delta$ times the critical density of the Universe, implying the three-dimensional virial mass is $M_{\Delta}=4\pi\Delta\rho_{c}r_{\Delta}^{3}/3$ (throughout the remainder of the article we use $\Delta=200$).  Constructing the lens models then requires the three-dimensional (deprojected) density distribution for each profile, as well as the three-dimensional mass distribution.  In general, these two equations can be inverted to find $r_{\Delta}$, and also the various parameters associated with the individual profiles (for example the velocity dispersion, $\sigma$, for the SIS profile).  The S\'ersic profile is a little more difficult to treat with this procedure as analytic forms of the deprojected density and mass functions do not exist.  As such, we use the analytic approximations given by \cite{prugniel97}, as well as empirical relations from \cite{graham06a}.  Details of the way in which we construct S\'ersic-law density distributions are provided in appendix \ref{PSapp}.

Following \cite{bacon06}, we use a flat $\Lambda$CDM cosmology with $\Omega_{m}=0.3$, $\Omega_{\Lambda}=0.7$ and $h=0.72$.  We place the lensing mass at a redshift of $z_{L}=0.35$ and the source at $z_{S}=0.8$ as these values correspond to $D_{LS}/D_{S}\simeq0.5$.  To compare the four density profiles, we use a lens mass of $M_{200}=10^{12}h^{-1}M_{\odot}$.  For the NFW profile, the concentration factor, $c$, which is the ratio of the virial radius to the scale radius (see section \ref{NFWsec}), is a function of the cosmology and the redshift of the lens which, for the above system, is evaluated to be $c=7.20$.  For the S\'ersic model we use empirically derived relations between the S\'ersic shape parameter, $n$, and the mass of the system: from \cite{graham06a}, their equation (12), we find the S\'ersic shape parameter for a galaxy of mass $M_{200}=10^{12}h^{-1}M_{\odot}$ is $n\simeq8.6$.

\begin{figure*}
\includegraphics[scale=0.6]{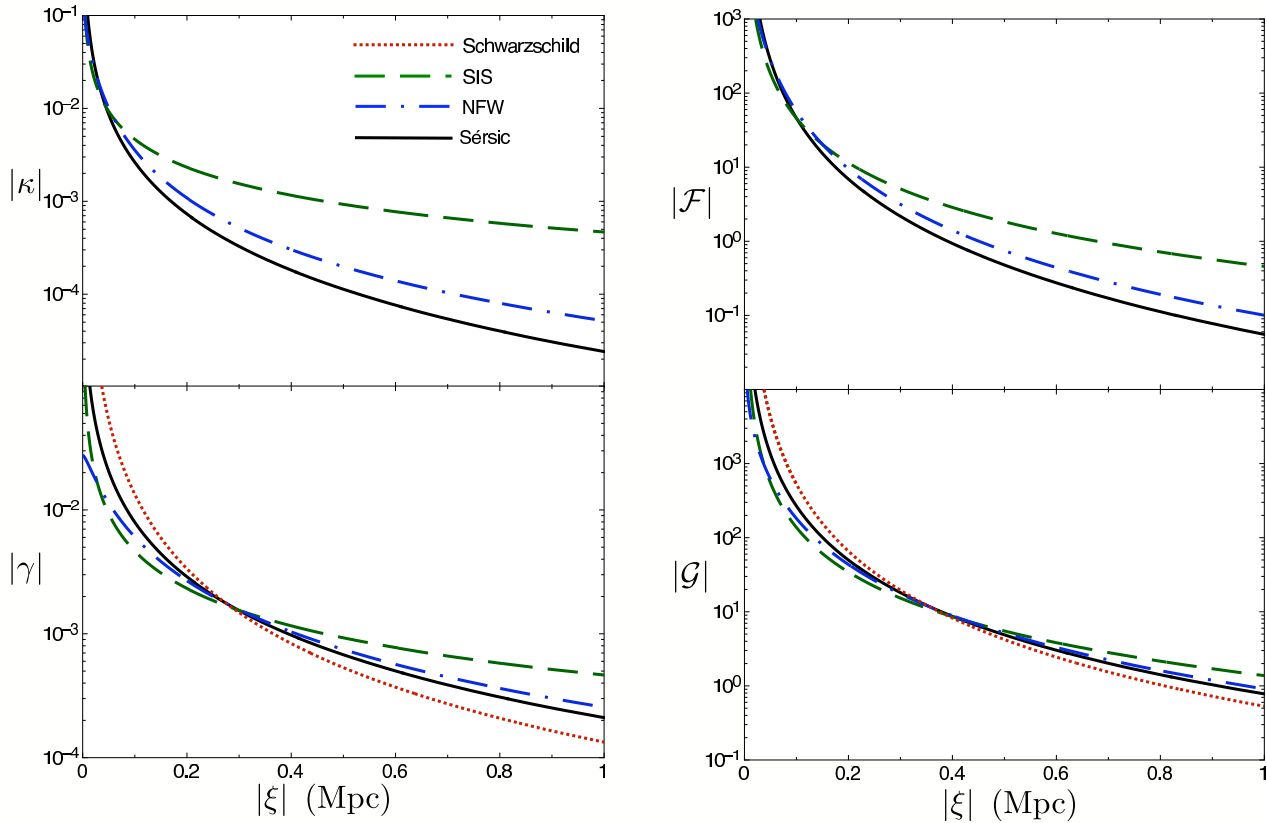}
\caption{Convergence, shear, first- and second-flexion for Schwarzschild lens (dotted red line), SIS (dashed green line), NFW (dash-dot blue line) and S\'ersic profile (thick black line).  The three-dimensional virial masses of each of these profiles is $10^{12}h^{-1}M_{\odot}$, with the lens and source placed at $z_{L}=0.35$ and $z_{S}=0.8$ respectively.  As an indicator of the weak lensing regime in these plots, the Einstein radius for the Schwarzschild lens is $\xi_{E}\simeq 11.5$ kpc.}
\label{allallMpc} 
\end{figure*}

Figure \ref{allallMpc} 
shows the lensing properties of the Schwarzschild lens (dotted red line), SIS (dashed green line), NFW (blue dashed-dot line) and S\'ersic (thick black line) mass distributions.  The Schwarzschild lens has an Einstein radius of $\xi_{E}\simeq11.5$ kpc, which implies that for impact parameters inside this radius the results presented are not in the weak lensing regime.  The most striking feature in these plots is that at large distances, the convergence and first-flexion for the SIS are significantly larger than for the NFW and S\'ersic profiles.  Bearing in mind that the convergence is linearly proportional to the surface density, this implies that at large distances the surface density of the SIS is also significantly larger than the NFW and S\'ersic profile.  This is consistent with the fact that the outer-logarithmic slope for NFW profiles is $-3$, while the slope of an isothermal sphere is $-2$.  Additionally, the slope of a S\'ersic-law profile depends on the specific shape parameter, $n$, and varies as a function of the distance from the centre of the profile, i.e. $|\xi|$.  Interestingly, whilst these features are evident in the convergence and the first-flexion, they are less apparent in the shear and second-flexion.



\begin{figure*}
\includegraphics[scale=0.6]{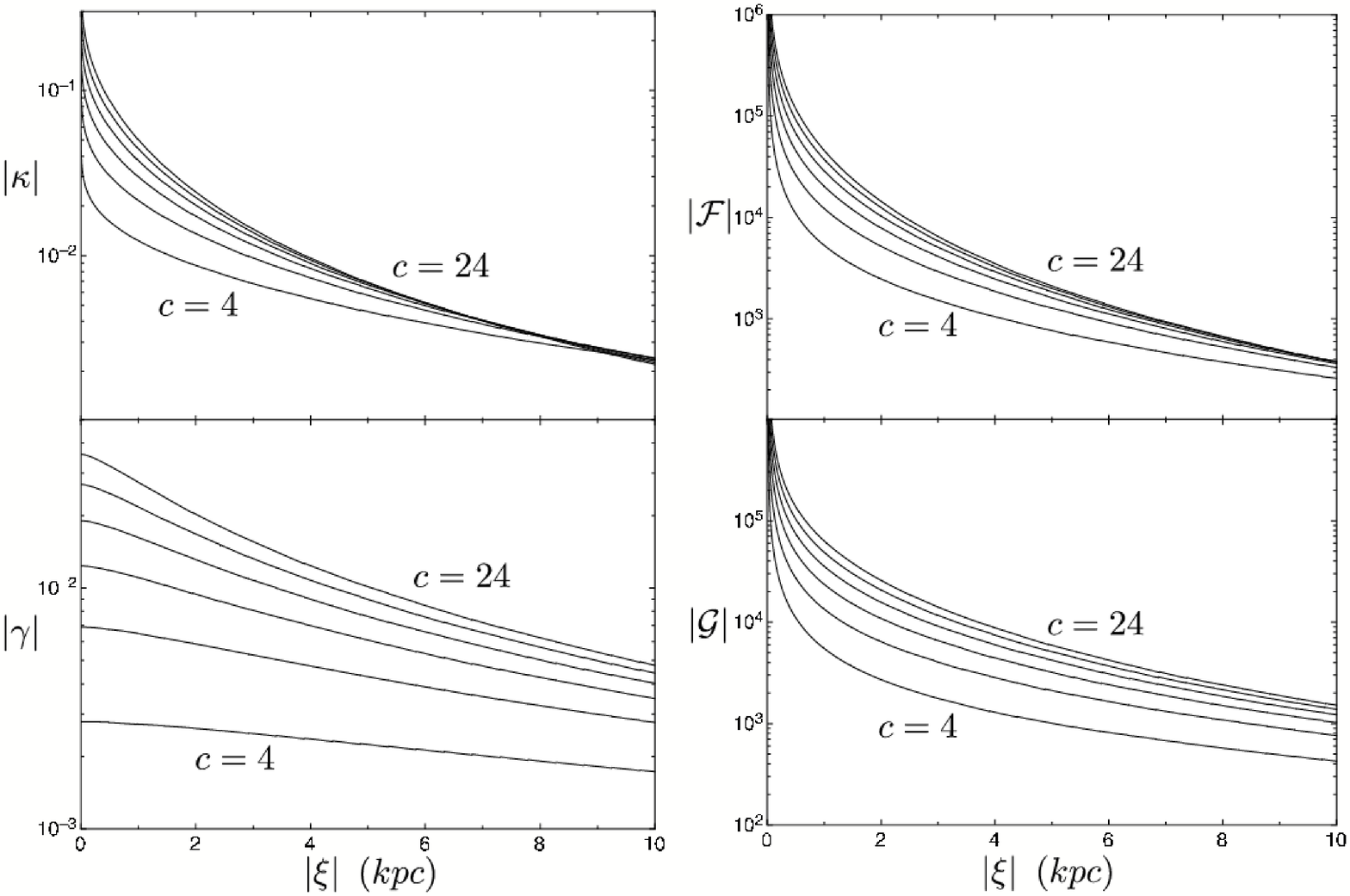}
\caption{Convergence, shear, first- and second-flexion for NFW profiles with ranging values of the concentration, $c$, with $M_{200}=10^{10}h^{-1}M_{\odot}$.  Note that a Schwarzschild lens of this mass at this distance has an Einstein radius of $\xi_{E}\simeq 1.2$ kpc, implying the weak lensing regime is somewhat beyond this limit.  The values of the concentration shown are $c=4,\,8,\,12,\,16,\,20,\,24$.  As the concentration increases, the curves converge on one another.  The shear and the second-flexion show distinctions between the various values of the concentration better than the convergence and first-flexion.}
\label{NFWvarycall} 
\end{figure*}

It is worth exploring the extent to which the above properties of the NFW and S\'ersic profiles depend on the concentration and shape parameters respectively.  The concentration parameter in the NFW profile, $c$, is defined as the ratio of the virial radius to the scale radius, which is a function of the specific cosmology.  Figure \ref{NFWvarycall} shows the effect of a varying concentration on the different lensing properties.  A lensing mass of $M_{200}=10^{10}h^{-1}M_{\odot}$ is used, with values of $c=4,\,8,\,12,\,16,\,20,\,24$.  For a Schwarzschild lens of this mass at these distances the Einstein radius is $\xi_{E}\simeq 1.2$ kpc, which again gives us a scale on which a weak lensing treatment is appropriate.

It is apparent from figure \ref{NFWvarycall} that the lensing properties of an NFW profile are not linearly effected by the concentration parameter.  Indeed as $c$ becomes larger, the lensing properties begin to converge, implying that at large $c$ the plots become indistinguishable.  Moreover, the effect of the concentration parameter is greater at small distances from the centre of mass of the lensing galaxy.  This is seen most pertinently in the convergence where at $|\xi|\sim10\xi_{E}$ the lines are indistinct.  It is also interesting to note that variations in the concentration parameter cause the shear and second-flexion to change significantly more than the convergence and first-flexion.  Therefore, given a series of lensed images, and assuming an NFW fit to the density profile, one can learn more about the specifics of the profile from the shear and the second-flexion than from the convergence and the first-flexion.  There is likely to be a degeneracy between the mass of the lensing galaxy and the concentration parameter if there are only a handful of images.  Our results imply that using both the shear and the second-flexion may be able to break this degeneracy so $c$ and $M$ can be obtained.

\begin{figure*}
\includegraphics[scale=0.6]{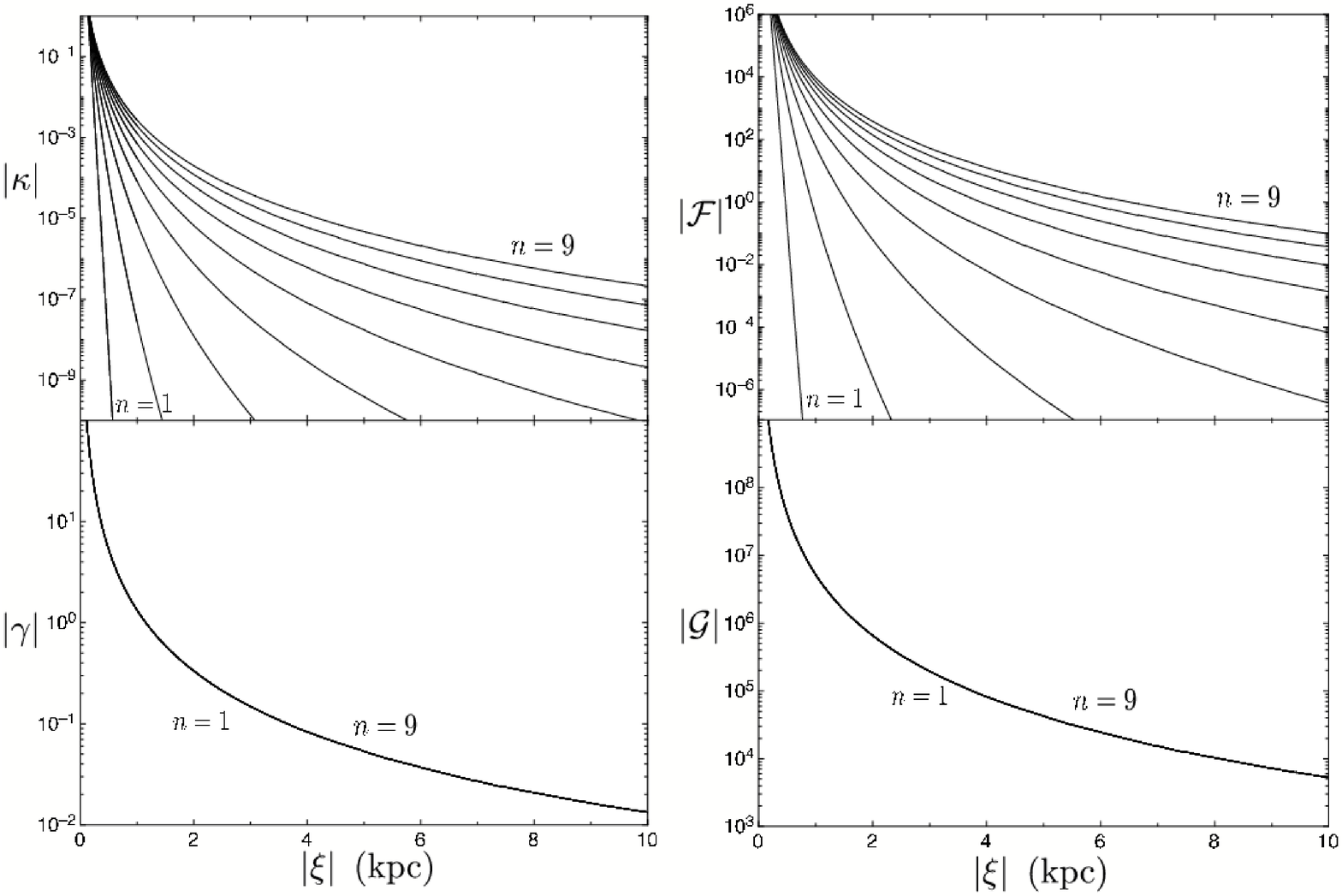}
\caption{Convergence, shear, first- and second-flexion for S\'ersic profiles with ranging values of the S\'ersic shape parameter, $n$, with $M_{200}=10^{10}h^{-1}M_{\odot}$.  As with figure \ref{NFWvarycall}, a Schwarzschild lens of this mass at this distance would have an Einstein radius of $\xi_{e}\simeq1.2$ kpc, implying the weak lensing regime is somewhat beyond this limit.  The shear and second-flexion vary extremely weakly with a change in $n$, whilst the convergence and first-flexion are more heavily dependent on this parameter.}
\label{Sersicvarynall}
\end{figure*}

Figure \ref{Sersicvarynall} shows the lensing properties of S\'ersic profiles for various values of the S\'ersic shape parameter, $n$.  The lensing mass is again $M_{200}=10^{10}h^{-1}M_{\odot}$, implying the point mass lens has $\xi_{E}\simeq1.2$ kpc, and we vary $1\le n\le9$.  
The striking feature of these plots is that the shear and second-flexion have significantly less dependence on the shape, $n$, of the S\'ersic profile than the convergence and the first-flexion.  
Indeed, zooming in on the vertical scale by an order of magnitude for the shear and second-flexion plots reveals that the dependence on $n$ can only be seen for $|\xi|\ll\xi_{E}$.
Moreover, the effect of the shape parameter on the convergence and first-flexion increases as one moves further from the source, although this may be difficult to detect observationally due to the relative size of the signal being significantly weaker at large separations from the lensing mass.    The fact that there is no dependence on the specific shape of the S\'ersic profile on the shear and second-flexion implies that these two properties can be used to derive the mass of the lensing object, whilst the convergence and first-flexion can then be used to derive $n$.  Weak gravitational lensing thus provides an independent method for deriving the masses of S\'ersic-law galaxies and clusters which only weakly depends on the specific shape of the profile, provided the correct lensing properties are utilised, i.e. the convergence and second-flexion.  

\section{Conclusion}\label{conc}
We have derived general equations governing the convergence, shear, first- and second-flexion for circularly symmetric gravitational lenses in terms of the surface density and projected mass of the lens.  We have shown that the components of the first-flexion are simply the directional derivatives of the surface density, while the second-flexion is a slightly more complicated function of the surface density, its gradient and also the projected mass distribution.  By applying the formalism to specific lens models, in particular a Schwarzschild lens, a singular isothermal sphere (SIS), Navarro-Frenk-White (NFW) profile and a S\'ersic profile, we have compared the signature each profile has on each of the lensing terms as a function of the radial impact parameter.  Whilst the NFW and SIS profiles have been compared previously, both at linear-order \citep{wright00} and also for flexion \citep{bacon06}, to the best of our knowledge this is the first time flexion for S\'ersic-law profiles have been presented.  In particular, we showed that the shear and second-flexion effects for a S\'ersic profile are systematically larger than for the NFW and SIS profiles (figure \ref{allallMpc}).  This implies that one must be careful about the specific matter distribution assumed when attempting to analytically reconstruct the total mass and mass profile of a circularly symmetric lens.  

An SIS profile is uniquely determined by designating the mass at a certain radius (e.g. virial mass).  However, both the NFW and S\'ersic models have extra parameters that determine their exact shape.  The concentration of the NFW profile is the ratio of the virial radius to the scale radius, which is a function of the particular cosmology being considered.  We showed that the convergence and first-flexion are relatively weakly dependent on the concentration parameter used, compared with the shear and the second-flexion (see figure \ref{NFWvarycall}).  The S\'ersic profile is parametrized by the S\'ersic shape parameter, $n$.  Counter to the NFW concentration parameter, the S\'ersic shape parameter is more heavily dependent on the convergence and first-flexion, whereas the shear and second-flexion vary little as $n$ changes (figure \ref{Sersicvarynall}).  These properties therefore provide the opportunity to directly measure both the mass of lensing galaxies as well as the specific shape of their density profiles.  Given a limited supply of images around a specific galaxy, higher-order lensing terms may provide the ability to give extra constraints on the individual profiles of the galaxies.  

The gravitational lensing systems explored in this article are idealized in that the projection of their mass distributions are circularly symmetric.  Moving beyond circular symmetry requires the numerical solution of the thin-lens gravitational lens equation.  
One is then free to study the effect of non-circular lens models by including anisotropies in the lensing galaxy, and also to look at the effect the size of the source has on the shape of the final image.  Whilst these applications will be extremely useful for determining the mass distributions of various galaxies using next-generation gravitational lensing surveys, the higher-order gravitational lensing effects will also be useful in studying the dark matter and dark energy content of the Universe.  Moreover, although the study of first-order gravitational lensing through N-body simulations has been around for some time \citep[see for example the recent review by][]{munshi08}, the study of flexion in these systems has scarcely been broached.  In this way, one can study the expected probability distributions for first- and second-flexion as a function of the specific cosmology.  The advantage of using flexion as opposed to linear gravitational lensing effects is that a systematic bias is taken out of the study as only one reasonable assumption about the source object is required -- that it is not intrinsically flexed.

\appendix
\section{Convergence, shear and flexion in angular coordinates}\label{anglesapp}

It is instructive to show the main equations from sections \ref{firstordersec} and \ref{flexionsec} expressed in angular coordinates, $(\theta_{1},\,\theta_{2})$, where $\xi_{i}=D_{L}\theta_{i}$.  From the definition of the projected mass function in terms of the surface density, equation (\ref{massdef}), implies that the projected mass expressed in angular coordinates is related to the mass in distance coordinates according to $M(|\xi|)=D_{L}^{2}M(|\theta|)$, where $|\theta|=\sqrt{\theta_{1}^{2}+\theta_{2}^{2}}$.  It is trivial to show that the expression for the convergence, equation (\ref{kappa}) goes unchanged, however the components of the shear, equations (\ref{gamma1}) and (\ref{gamma2}), written in angular coordinates become
\begin{eqnarray}
	\gamma_{1}(\theta_{1},\theta_{2})&=&\frac{\theta_{1}^{2}-\theta_{2}^{2}}{\pi\Sigma_{cr}|\theta|^{4}}\left(\pi\Sigma|\theta|^{2}-M\right)\\
	\gamma_{2}(\theta_{1},\theta_{2})&=&\frac{2\theta_{1}\theta_{2}}{\pi\Sigma_{cr}|\theta|^{4}}\left(\pi\Sigma|\theta|^{2}-M\right),
\end{eqnarray}
where for the remainder of this appendix $\Sigma=\Sigma(|\theta|)$ and $M=M(|\theta|)$.  The magnitude of the total shear is
\begin{equation}
	\left|\gamma(\theta_{1},\theta_{2})\right|=\frac{1}{\pi\Sigma_{cr}|\theta|^{2}}\left(\pi\Sigma|\theta|^{2}-M\right).
\end{equation}
The components of the first-flexion, equations (\ref{F1gen}) and (\ref{F2gen}), are given in angular coordinates as
\begin{eqnarray}
	\F_{1}(\theta_{1},\theta_{2})&=&\frac{1}{\Sigma_{cr}}\frac{\partial\Sigma}{\partial\theta_{1}},\\
	\F_{2}(\theta_{1},\theta_{2})&=&\frac{1}{\Sigma_{cr}}\frac{\partial\Sigma}{\partial\theta_{2}},
\end{eqnarray}
implying the magnitude of the first-flexion is
\begin{equation}
	\left|\F\left(\theta_{1},\theta_{2}\right)\right|=\frac{1}{\Sigma_{cr}}\frac{\da\Sigma}{\da|\theta|}.
\end{equation}
Where previously the flexion terms were expressed in units of distance, one can now see first-flexion has units of (angle)$^{-1}$.  It is trivial to see that the following three expressions governing the second-flexion in terms of angular coordinates are also expressed in these same units.
\begin{eqnarray}
	\G_{1}\left(\theta_{1},\theta_{2}\right)&=&\frac{\theta_{1}\left(\theta_{1}^{2}-3\theta_{2}^{2}\right)}{\pi\Sigma_{cr}|\theta|^{6}}\left[\pi\frac{d\Sigma}{d|\theta|}|\theta|^{3}-4\pi\Sigma|\theta|^{2}+4M\right],\\
	\G_{2}\left(\theta_{1},\theta_{2}\right)&=&\frac{\theta_{2}\left(3\theta_{1}^{2}-\theta_{2}^{2}\right)}{\pi\Sigma_{cr}|\theta|^{6}}\left[\pi\frac{d\Sigma}{d|\theta|}|\theta|^{3}-4\pi\Sigma|\theta|^{2}+4M\right],\\
	\left|\G\left(\theta_{1},\theta_{2}\right)\right|&=&\frac{1}{\Sigma_{cr}}\left(\frac{d\Sigma}{d|\theta|}-\frac{4\Sigma}{|\theta|}+\frac{4M}{\pi|\theta|^{3}}\right).
\end{eqnarray}

\section{Creating Galaxies with S\'ersic Profiles}\label{PSapp}
The S\'ersic profile is defined in terms of the projected surface density, equation (\ref{Sersic}).  However, to compare the lensing effects of the S\'ersic profile with other density distributions, one requires the deprojected form of the density and mass distributions (see the discussion in section \ref{comparesec}).  Whilst an analytic form of the deprojected S\'ersic profile is not available, an analytic approximation has been provided by\cite{prugniel97}, and has further been explored in detail by \cite{merritt06,graham06,graham06a}.  

\cite{prugniel97} showed that the three-dimensional density distribution associated with the surface density of the S\'ersic profile, given by equation (\ref{Sersic}), can be approximated as 
\begin{equation}
	\rho(r)=\rho_{e}\left(\frac{r}{\xi_{e}}\right)^{-p}\exp\left\{-b_{n}\left[\left(\frac{r}{\xi_{e}}\right)^{1/n}-1\right]\right\}.\label{PS}
\end{equation}
Here, $\rho_{e}$ is the three-dimensional density at the effective (projected) radius $\xi_{e}$.  The function $p=p(n)$ is utilised to ensure that the projection of equation (\ref{PS}) relates as closely as possible to the projected S\'ersic profile, i.e. equation (\ref{Sersic}), for the range $0.6\le n\le10$.  This was first given by \cite{limaneto99} as $p=1.0-0.6097/n+0.05463/n^{2}$, and a goodness of fit is shown in \cite{merritt06}.  Integrating the three-dimensional density distribution over the volume gives the three-dimensional mass distribution,
\begin{equation}
	M_{3D}(r)=4\pi n\xi_{e}^{3}\rho_{e}{\rm e}^{b_{n}}b_{n}^{-\left(3-p\right)n}\Gamma\left[\left(3-p\right)n,\,b_{n}\left(\frac{r}{\xi_{e}}\right)^{1/n}\right].\label{Sersic3dmass}
\end{equation}

\cite{cardone04} first looked at gravitational lensing for a S\'ersic profile.  He discussed the need to reduce the parameter space of the system in order to build a graviational lens from a S\'ersic model.  To that end, he used an empirical relation governing the deprojected effective radius, the central surface brightness and the S\'ersic shape parameter, $n$ (see equation A.4 and A.5 of \cite{cardone04}).  It is possible for us to also use this relation, and subsequently convert the central surface brightness into a density by invoking more empirical relations and also assuming a mass-to-light ratio.  Essentially, this procedure has already been completed for the Prugniel \& Simien model by \cite{graham06a} [their equations (13) and (14)];
\begin{equation}
	\log_{10}\rho_{e}=k-2.5\log_{10}\xi_{e}.\label{rhoRe}
\end{equation}
Here, $\xi_{e}$ is in units of kiloparsecs, $\rho_{e}$ is in solar masses per cubic parsec and $k$ is a constant which is $0.5$ for luminous elliptical galaxies and galaxy-sized dark matter halos (with $\log R_{e}\gtrsim0.5$) and $2.5$ for cluster-sized dark matter halos (with $\log R_{e}\gtrsim1.5$).  Finally, the three-dimensional density at the effective (projected) radius is related to the two-dimensional surface density at $\xi_{e}$ by
\begin{equation}
	\rho_{e}=\Sigma_{e}b_{n}^{\left(1-p\right)n}\frac{\Gamma\left(2n\right)}{2\xi_{e}\Gamma\left[\left(3-p\right)n\right]}.\label{rhoeSigmae}
\end{equation}

As mentioned in section \ref{comparesec}, to compare profiles we specify the virial mass of the system, $M_{\Delta}$,
implying we know the virial radius, $r_{\Delta}$.  Substituting this into equation (\ref{Sersic3dmass}), together with equation (\ref{rhoRe}) implies we have an equation for $\xi_{e}$ as a function of the S\'ersic shape parameter, $n$.  This equation is not analytically invertible
due to the presence of the incomplete gamma function, however it can be solved numerically for given values of $n$.  Therefore, once this equation is solved, we know $\xi_{e}$ and hence $\rho_{e}$, which can both be substituted into equation (\ref{rhoeSigmae}) to give $\Sigma_{e}$.  The projected S\'ersic profile (\ref{Sersic}) can finally be evaluated, along with its various derivatives and also the two-dimensional mass, implying all of the lensing quantities can be evaluated.

\section*{acknowledgments}
We thank Alister Graham for helpful comments regarding the S\'ersic profile and the Prugniel \& Simien model.  We also thank the referee for their extremely thorough and insightful review of the original manuscript.  This research was supported under the Australian Research Councils Discovery Projects funding scheme (project number DP0665574).

\bibliography{Flexion.bib}
\bibliographystyle{mn2e}
\bsp
\label{lastpage}

\end{document}